\documentclass[acmsmall,authorversion]{acmart}

\usepackage{listings}
\usepackage{algorithm}
\usepackage{algpseudocode}

\usepackage{graphicx}

\usepackage{lscape}
\usepackage{rotating}
\usepackage{pdflscape}
\usepackage{placeins}

\AtBeginDocument{%
  \providecommand\BibTeX{{%
    \normalfont B\kern-0.5em{\scshape i\kern-0.25em b}\kern-0.8em\TeX}}}

\copyrightyear{2023}
\acmYear{2023}
\setcopyright{acmlicensed}\acmConference[WWW '23 Companion]{Companion Proceedings of the ACM Web Conference 2023}{April 30-May 4, 2023}{Austin, TX, USA}
\acmBooktitle{Companion Proceedings of the ACM Web Conference 2023 (WWW '23 Companion), April 30-May 4, 2023, Austin, TX, USA}
\acmPrice{15.00}
\acmDOI{10.1145/3543873.3587601}
\acmISBN{978-1-4503-9419-2/23/04}

\begin{document}

\title[Application of an ontology for model card for linking machine learning information from biomedical research]{Application of an ontology for model cards to generate computable artifacts for linking machine learning information from biomedical research}


\author{Muhammad ``Tuan'' Amith}
\affiliation{%
	\institution{University of North Texas}
	\streetaddress{3940 North Elm, Suite E292}
	\city{Denton}
	\country{USA}}
\email{muhammad.amith@unt.edu}

\author{Licong Cui}
\affiliation{%
	\institution{The University of Texas Health Science Center at Houston}
	\streetaddress{7000 Fannin St, Suite 600}
	\city{Houston}
	\country{USA}}
\email{licong.cui@uth.tmc.edu}

\author{Kirk Roberts}
\affiliation{%
	\institution{The University of Texas Health Science Center at Houston}
	\streetaddress{7000 Fannin St, Suite 600}
	\city{Houston}
	\country{USA}}
\email{kirk.roberts@uth.tmc.edu}

\author{Cui Tao}
\authornote{corresponding author}
\affiliation{%
	\institution{The University of Texas Health Science Center at Houston}
	\streetaddress{7000 Fannin St, Suite 600}
	\city{Houston}
	\country{USA}}
\email{cui.tao@uth.tmc.edu}

\renewcommand{\shortauthors}{Amith, et al.}

\begin{abstract}
  Model card reports provide a transparent description of machine learning models which includes information about their evaluation, limitations, intended use, etc. Federal health agencies have expressed an interest in model cards report for research studies using machine-learning based AI. Previously, we have developed an ontology model for model card reports to structure and formalize these reports. In this paper, we demonstrate a Java-based library (OWL API, FaCT++) that leverages our ontology to publish computable model card reports. We discuss future directions and other use cases that highlight applicability and feasibility of ontology-driven systems to support FAIR challenges.
\end{abstract}




\maketitle

\section{Introduction}

The National Institutes of Health has expressed interest in applying FAIR principles to the growing number of biomedical research involving machine learning-based artificial intelligence (AI) resources \cite{national_institutes_of_health_bridge_2021}. In addition, the initiative also involves disclosing some transparent ethical information for these AI models. To accomplish this, there needs to be standards, software, and datasets to support FAIR principles \cite{wilkinson_fair_2016} and accommodating the aforementioned ethics. One of the suggested tools are \textit{model card reports} which are static documents that outline features and associated information about machine learning models \cite{mitchell_model_2019}.

Model card reports were introduced as a type of report to accompany the release of a machine learning model to document various aspects of the model. This short document of one to two pages would detail performance information, limitations, intended uses and information about the evaluation of the datasets used to train and test the model. This form of documentation intends to help standardize the practice of transparency for machine learning models for stakeholders.

In our previous work \cite{amith_toward_2022}, we developed an ontology (Model Card Report Ontology) to represent a model card report with the intent that the ontology can be used as a framework for computable digital representation of the model card. This was demonstrated by translating samples of model card report from experimental bioinformatics studies to show how a static model card report would map to our ontology framework. One of the challenging tasks of producing this computable linked version of the model card report was the arduous effort in using Prot\'eg\'e authoring tool \cite{musen_protege_2015} to encode the model card information. This is also compounded that novice users would find difficulty in using Prot\'eg\'e \cite{Rector2004}.

\begin{figure*}[h!]
	\centering
	\includegraphics[width=0.8\linewidth]{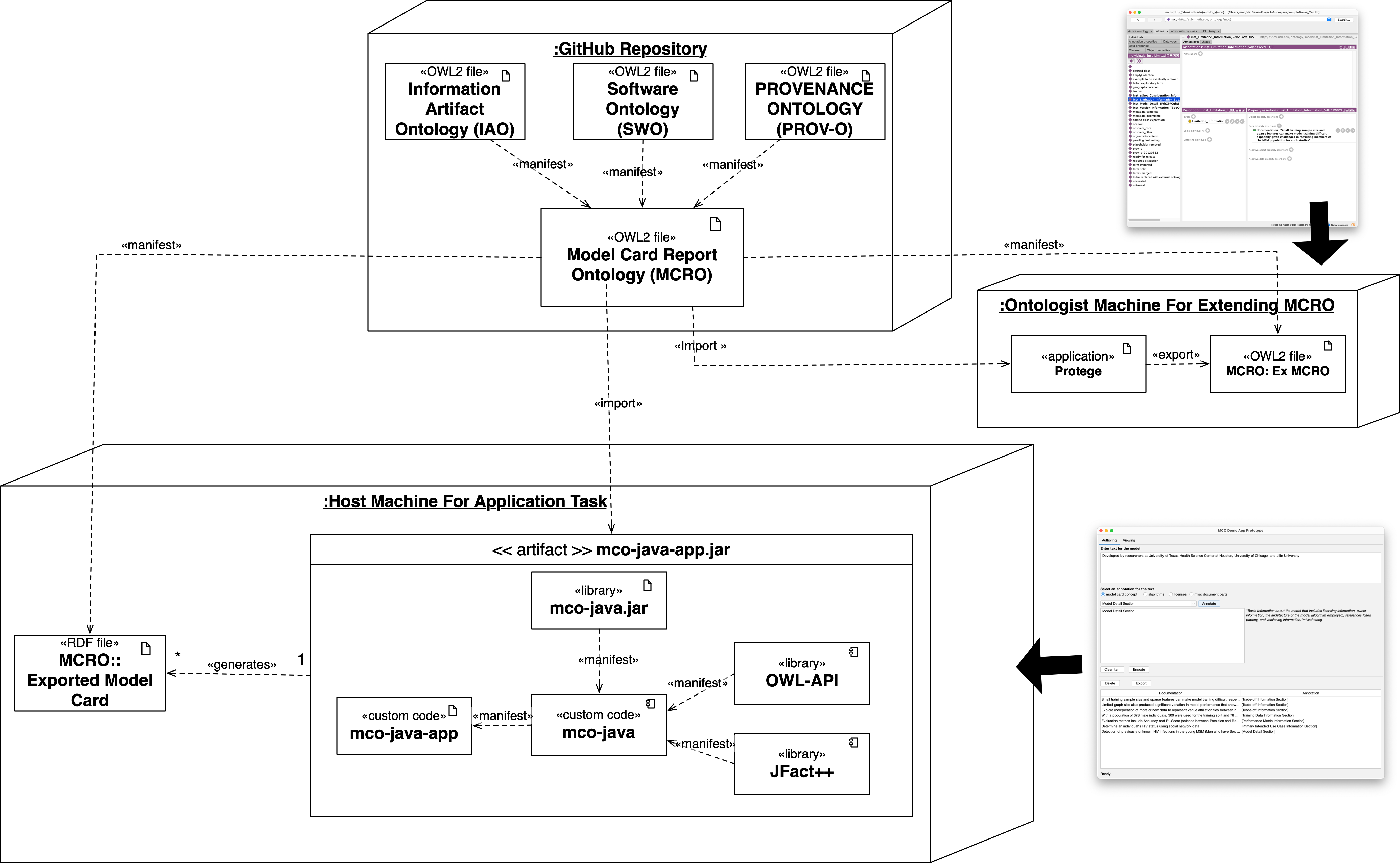}
	\caption{Overall architecture of a deployed Model Card Report Ontology for extended enrichment for use by an ontologist and for utilization for the publishing application engine using a Java wrapper library and front end.}
	\label{fig:architecture}
\end{figure*}

This work aims to circumvent the challenges and make the process of generating an ontology-based model card report accessible. We also introduce an ontology-driven publishing engine that would be part of an information processing pipeline to produce ontology-based model card reports for distribution and sharing. The goal is to demonstrate the applicability of ontologies in document engineering, specifically to show the mechanism of the engine using Semantic Web technology and also to formalize linked model cards using an ontology-based method.

We propose that our automated publishing engine can semantically link model card content to an ontology-based artifact that can be shared and reused. We will use the aforementioned Model Card Report Ontology (MCRO)\cite{amith_model_2021,amith_toward_2022} and various software API libraries that will enable the production of dynamic model card artifacts. The generation process will be illustrated by a user interface to show the accessibility of the library to author dynamic model card reports for digital distribution.

In this paper, we describe the system architecture and how the machine-based reasoner is able to link the content of a model card report together and output the report in a computable, machine readable format. We also describe our vision on how future natural language processing methods will integrate with the engine, as a future direction.

\section{Methods}

\subsection{Model Card Report Ontology}
OWL-based ontologies bore out of Tim Berners-Lee's vision of the Semantic Web \cite{berners2001semantic}. The Semantic Web entails a web of linked data that would be an evolution of the presentation-modality of the modern web. Part of the technology stack of the Semantic Web include ontologies to make the linkages of data possible. Semantics infused in the linked data enables machines to utilize the web of data for software agents to aggregate and collate data for intelligent reasoning. This is supported through the semantics provided through these ontologies. The Linked Open Data Cloud \cite{andrejs_abele_linking_2017} is one incremental example toward the Semantic Web vision.

As noted before, we developed an ontology artifact called the Model Card Report Ontology (MCRO), encoded with the Web Ontology Language (OWL2) \cite{w3c_owl_working_group_and_others_owl_2012}. This ontology formally represents the core structure of the model card report adopted from the original model card design \cite{mitchell_model_2019,google_inc_model_2021}. In our previous work, we have demonstrated the utility of the ontology as a scaffold for linking instance data of a model card in the form of an RDF (Resource Description Framework) document. This approaches enables the sharing and linking of for potential aggregation and analysis, thereby enhancing the functionality of the model card report beyond a static document. MCRO is encoded in the Web Ontology Language (OWL2) \cite{w3c_owl_working_group_and_others_owl_2012} for consumption, and is available for public release at GitHub for utilization \cite{amith_model_2021}.

\begin{figure*}[h!]
	\centering
	\includegraphics[width=1\linewidth]{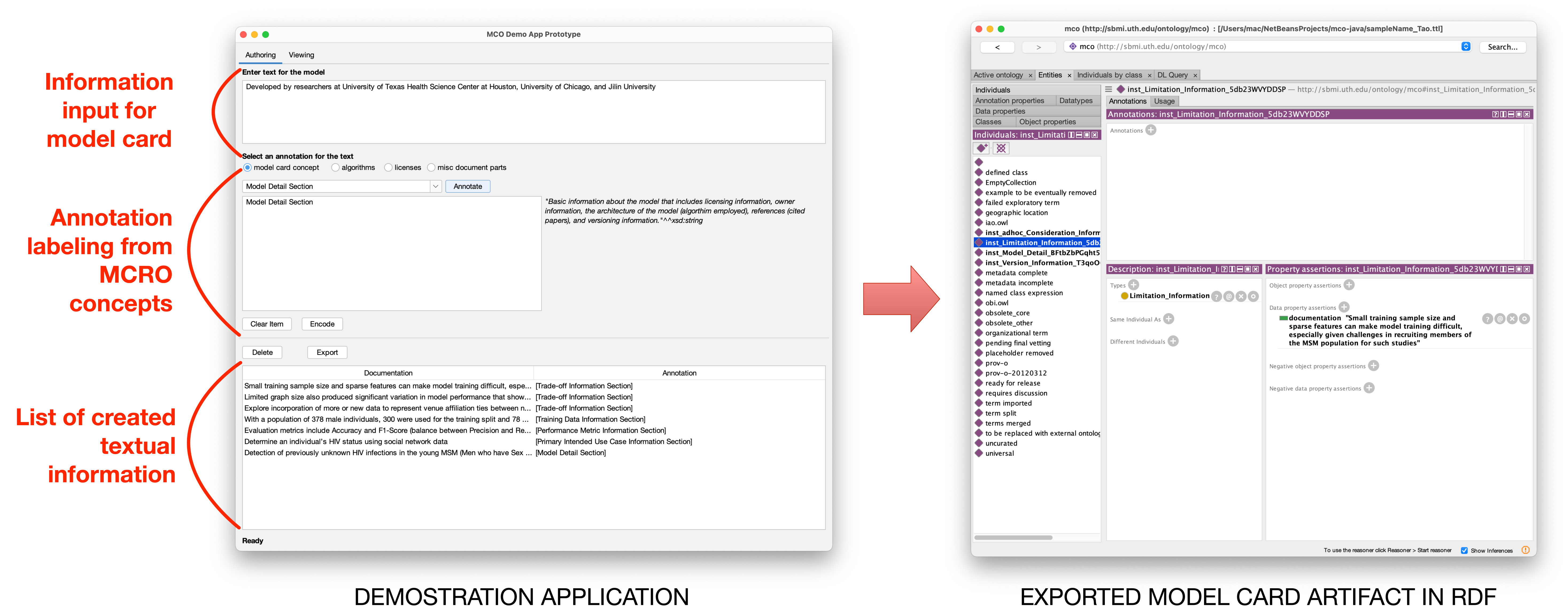}
	\caption{Screen shot of the application publishing engine (left) using the MCRO that can export an RDF-based model card report (viewed through Prot\'eg\'e, right)}
	\label{fig:export}
\end{figure*}

\subsection{Architecture Deployment}

In Figure \ref{fig:architecture}, we describe an architecture deployment of probable software system that leverages the MCRO. The ontology is provided through the aforementioned public repository where it can be imported for additional application tasks or further enriched with another ontology artifact.  

For the application perspective, we developed a Java-based library that utilizes  OWL API \cite{horridge2011owl} and FaCT++ \cite{tsarkov_fact_2006} APIs. This library enables the host's machine to import the ontology model from the web and utilizes the semantic definition (e.g., via Description Logic) for reasoning to format and generate a model card report. As a wrapper library, external tools can interface with the ontology model to add or link data to the ontology model structure. 

We developed a Java Swing front-end application to demonstrate the practical application of our ontology-driven library. Using this library, we were able to generate a linked model card report that can be utilized as a computable artifact (e.g., RDF, JSON, etc.). The artifact can be shared and distributed, and later archived for querying and analytical purposes. Figure \ref{fig:export} shows the application front end. The operationalization of the MCRO ontology essentially involves linking the textual data and information to the structure of the report document of the model card - framed by the MCRO. The experimental interface is designed to annotate each piece of information in the model card with categories (classes) with respect to the MCRO model, which also includes concepts from Software Ontology \cite{malone_software_2014} and the Information Artifact Ontology \cite{ceusters_information_2012} for additional relevant classes. In addition, the experimental interface demonstrates export and generation of the computable artifact of the model card report (see Figure \ref{fig:export})

\subsection{Managing Model Card Information using Deductive Reasoning from Description Logic}

One of the features of the Web Ontology Language (OWL2) is the ability to leverage the expressive semantics within the syntax to preform machine-based deductive reasoning. OWL2 implements Description Logic, a type representational logic utilizing axioms (instead of frames),  that involves \textit{TBox} (the schema to define the constraints of the ontology), \textit{ABox} (the data instantiated by the TBox), and \textit{RBox} (roles of the relationship between the data in the ontology) \cite{krotzsch_description_2012}.  Using MCRO, we are able to utilize inferencing to handle the production of an ontology-based model card report from the individual text information and derive the types of explicit and implicit concepts encoded in the ontology structure. 

We described the basic process in how our proposed software will generate a linked model card report in an ontology-based format (See Algorithm \ref{alg:dl}). During this process, each  text in the model card is annotated with its corresponding concept class (Algorithm \ref{alg:dl}, line \ref{alg:dl:for}). The associated class of the instance data is then  used to find its meronymy relationship (i.e., \textit{part of}), which is achieved using Description Logic (see Algorithm \ref{alg:dl}, line \ref{alg:dl:meronymy}). From the child class ($c_{child}$), we retrieve the instance data ($i_{child}$) of that child class, and then link that instance data to the instance ($i_{parent}$) of the parent class. This linked pair is then added to the ontology model.

\begin{algorithm}[h!]
	\caption{Basic procedure to use description logic to find and link instance data to the ontology}\label{alg:dl}
	\begin{algorithmic}[1]
		
		\Require MCO $=$ $\{C_n\},$ where $C_n = \{(c_n, i_n)\}$ is annotated model card text
		\State MCO $\gets$ in-memory model card ontology data
		\State DLQ $\gets$ description logic query reasoner
		\State OM $\gets$ ontology manager functions
		\For {$c \in \{C_n\}$} \label{alg:dl:for}
		\State $i_{parent}$ $\gets$ OM.class\_instance($c$)
		\State \label{alg:dl:meronymy} $c_{child}$ $\gets$ DLQ.meronymy\_of($c$)
		\State $i_{child}$ $\gets$ OM.class\_instance($c_{child}$)
		\State $L_{pair}$ $\gets$ OM.link($i_{parent}$,$i_{child}$)
		\State $MCO$ $\gets$ OM.add\_to\_ontology(MCO, $L_{pair}$)
		\EndFor
		
		\Ensure MCO $=$ $\{L_n, C_n\},$ where $L\_n$ is a linked pair
		
	\end{algorithmic}
\end{algorithm}

\begin{figure*}[h!]
	\centering
	\includegraphics[width=0.9\linewidth]{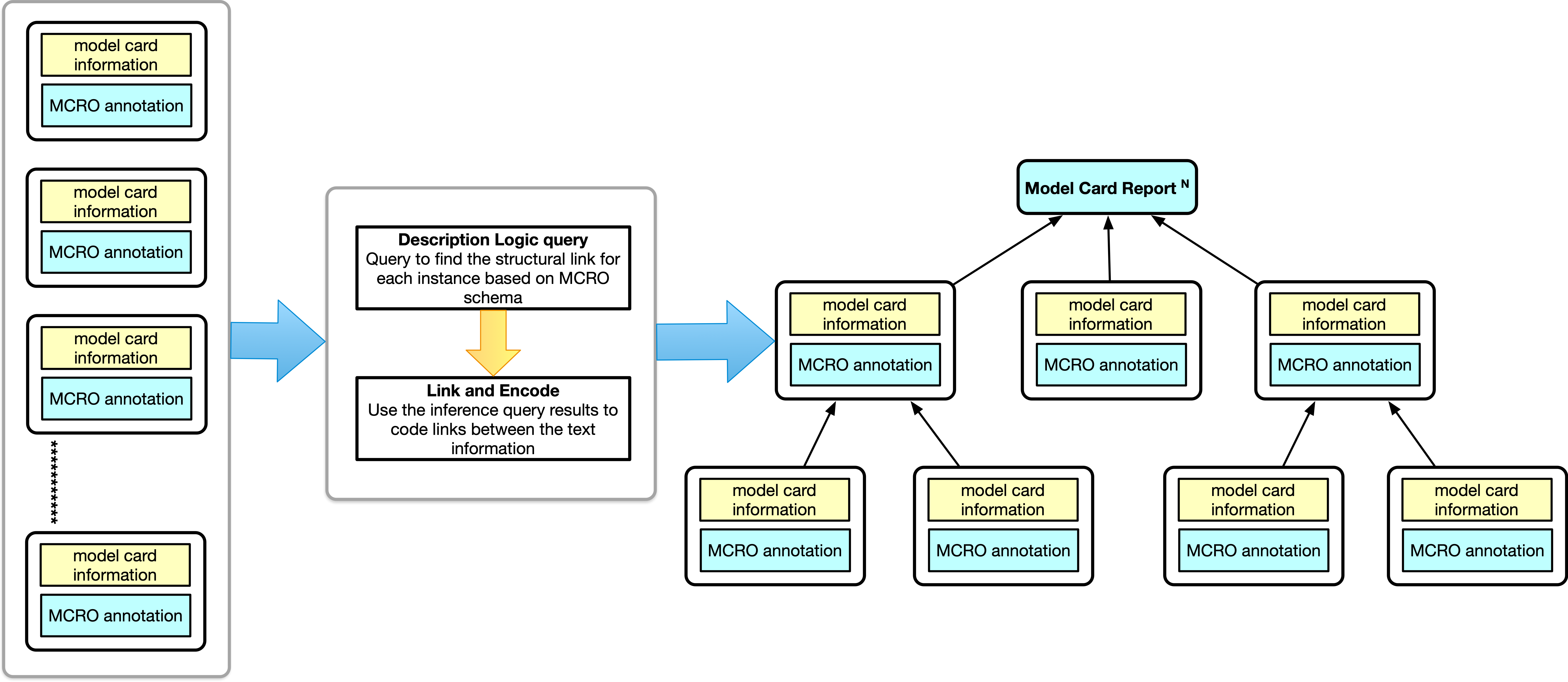}
	\caption{Abstract account of the underlying algorithm utilizing Description Logic reasoning to encode the links between the model card information from the application.}
	\label{fig:dl}
\end{figure*}

In Figure \ref{fig:dl}, we show an abstract visual that demonstrates how textual information generated from the application interface can be linked to  the independent sections of the model card report through inference using  Description Logic queries (``DL query''). Within MCRO, each concept corresponding to a model card section has a semantic definition of its link to a part of the model card document representation. Based off of the class definition, the query infers the link of each text instance and then the publishing engine encodes that link to the in-memory ontology. After all of the links are identified for each instance and encoded, the ontology-based model card is formed and ready for export in RDF, OWL, Turtle, or JSON.

\section{Results}

We created a prototype library (.jar) using Java 11 and several software API libraries - OWL API, Java implementation of FaCT++. As described earlier, the library retrieves the the Model Card Report Ontology online and uses it as a schema to represent the structure of a model card report. The library is hosted on our Github repository, https://github.com/UTHealth-Ontology/MCRO-Software, alongside a modern graphical user interface (GUI) to illustrate the accessibility of the prototype software package. 

\subsection{Integration with Demonstration Software}

To show the possibility of an ontology-based software engine integrated with software systems for publishing model cards, we developed a rudimentary prototype software application that leverages the software library. The sample application is available at our GitHub repository, and a downloadable cross-platform executable Java file is also hosted for testing. The sample software was developed in Java using modern Swing libraries. Using the interface, the software library exposes several functions that enable this tool (or any software system) to interact and manage data that is structured by the Model Card Report Ontology. We used a model card example from our previous work \footnote{ https://github.com/UTHealth-Ontology/MCRO/blob/main/samples/mc2.md} to walk through the basic usage of the software library through the interface.

\begin{figure}[h!]
	\centering
	\includegraphics[width=1\linewidth]{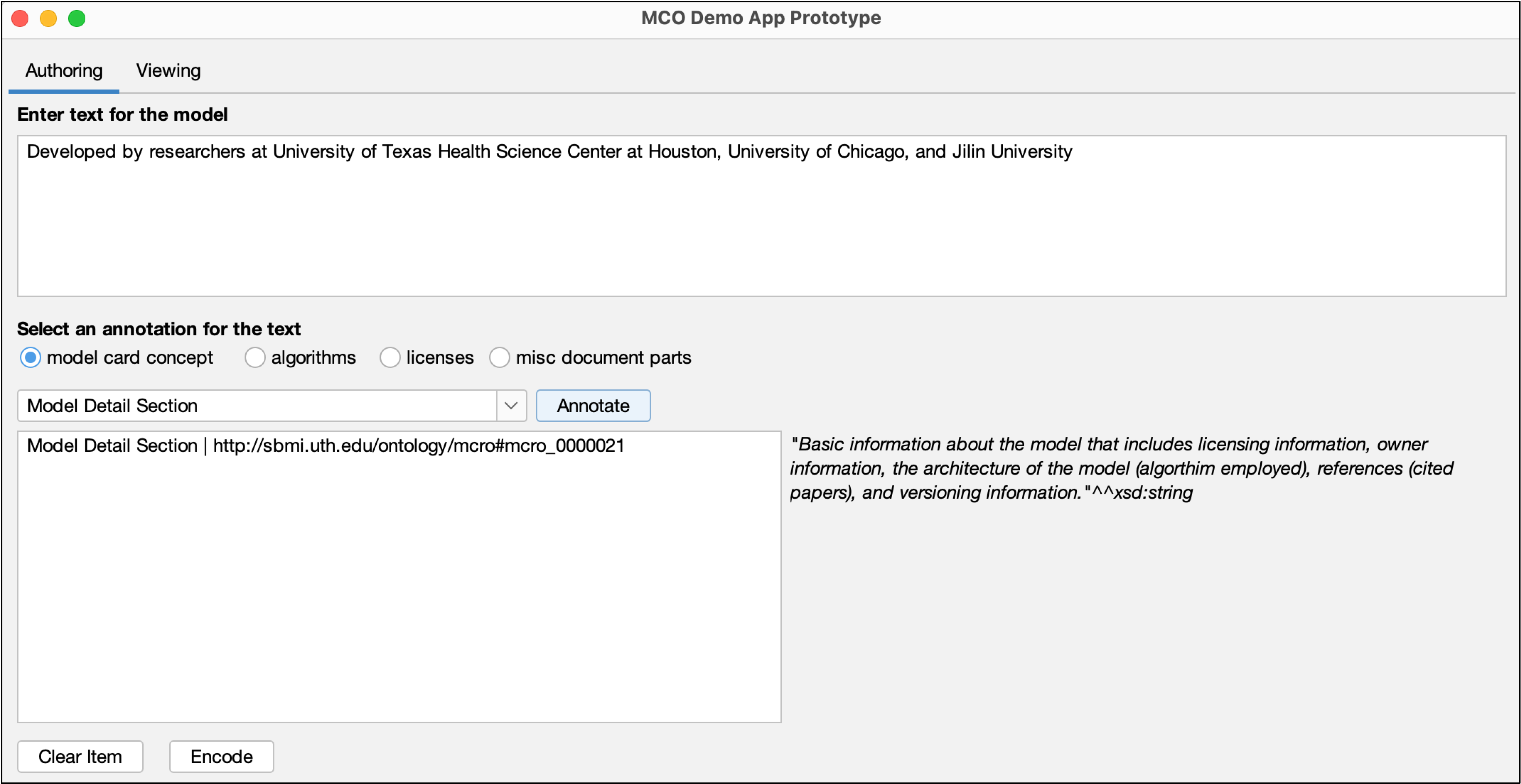}
	\caption{A section of the demonstration application for model card information to be inputted and annotated from concepts derived from the Model Card Report Ontology.}
	\label{fig:text-app}
\end{figure}

In Figure \ref{fig:text-app}, a prospective user inputs information about a machine learning model for annotation and labeling using the Model Card Report Ontology  and its linked auxiliary ontologies - Software Ontology, Information Artifact Ontology, etc. The aforementioned ontologies reside on their respective authors repositories through their Uniform Resource Identifiers (URIs). Therefore, any updates made to the ontologies would be available to the software library for use.  

\begin{figure}[h!]
	\centering
	\includegraphics[width=1\linewidth]{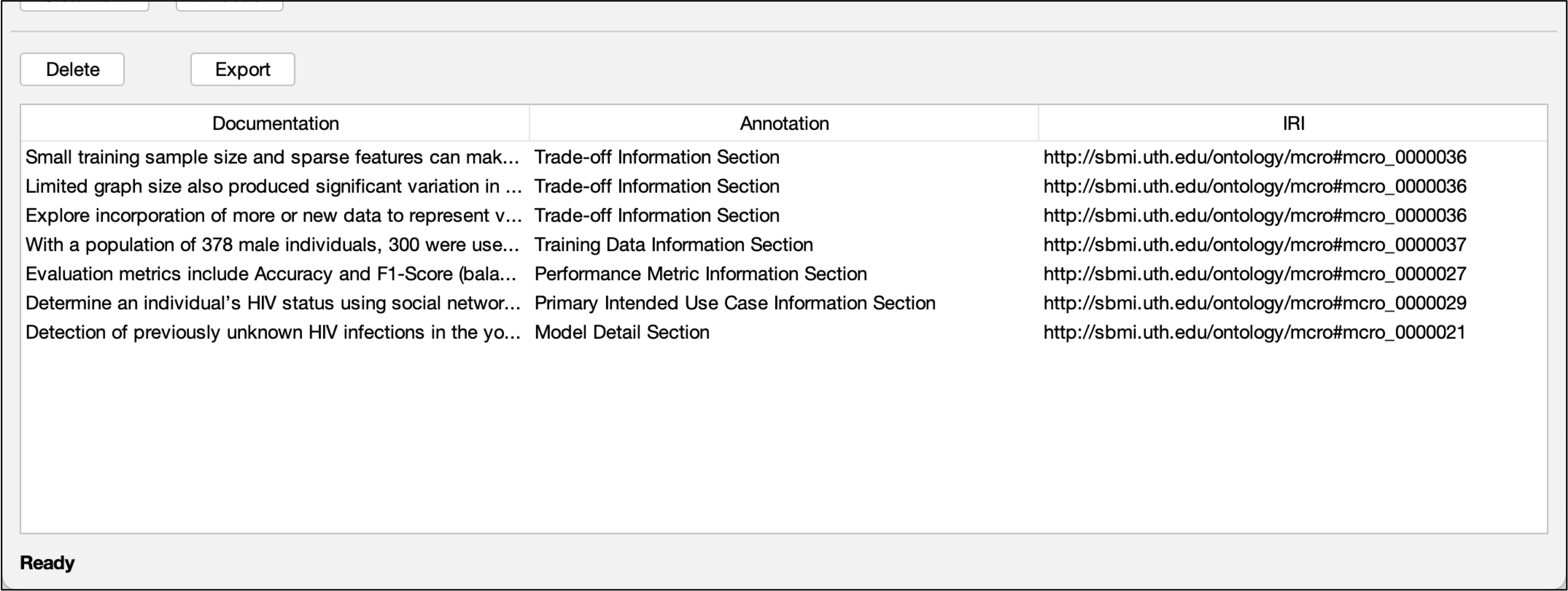}
	\caption{A part of the interface showing a list of annotated model card informational text, along with the class annotation's identifier}
	\label{fig:collection-app}
\end{figure}

After an atomic piece of information about the machine learning model is entered, the user can associate a category or class to the textual information. On Figure \ref{fig:text-app}, categories from the Model Card Report Ontology is presented to the user to retrieve a list of possible annotations from each category. The library retrieves a list of all subclasses for the model card concepts, algorithms and licenses (linked from the Software Ontology), and various document parts defined from the Information Artifact Ontology. The library also retrieves any linked annotation, like RDF comments. In the same figure, a description linked to the concept is displayed to help the user. Clicking "Annotate" associates a class concept from MCRO to the model card information. Figure \ref{fig:collection-app} shows (when the user clicks "Encode", see Figure \ref{fig:text-app}) the model card information that is attached to associated annotation(s).

\begin{figure}[h!]
	\centering
	\includegraphics[width=1\linewidth]{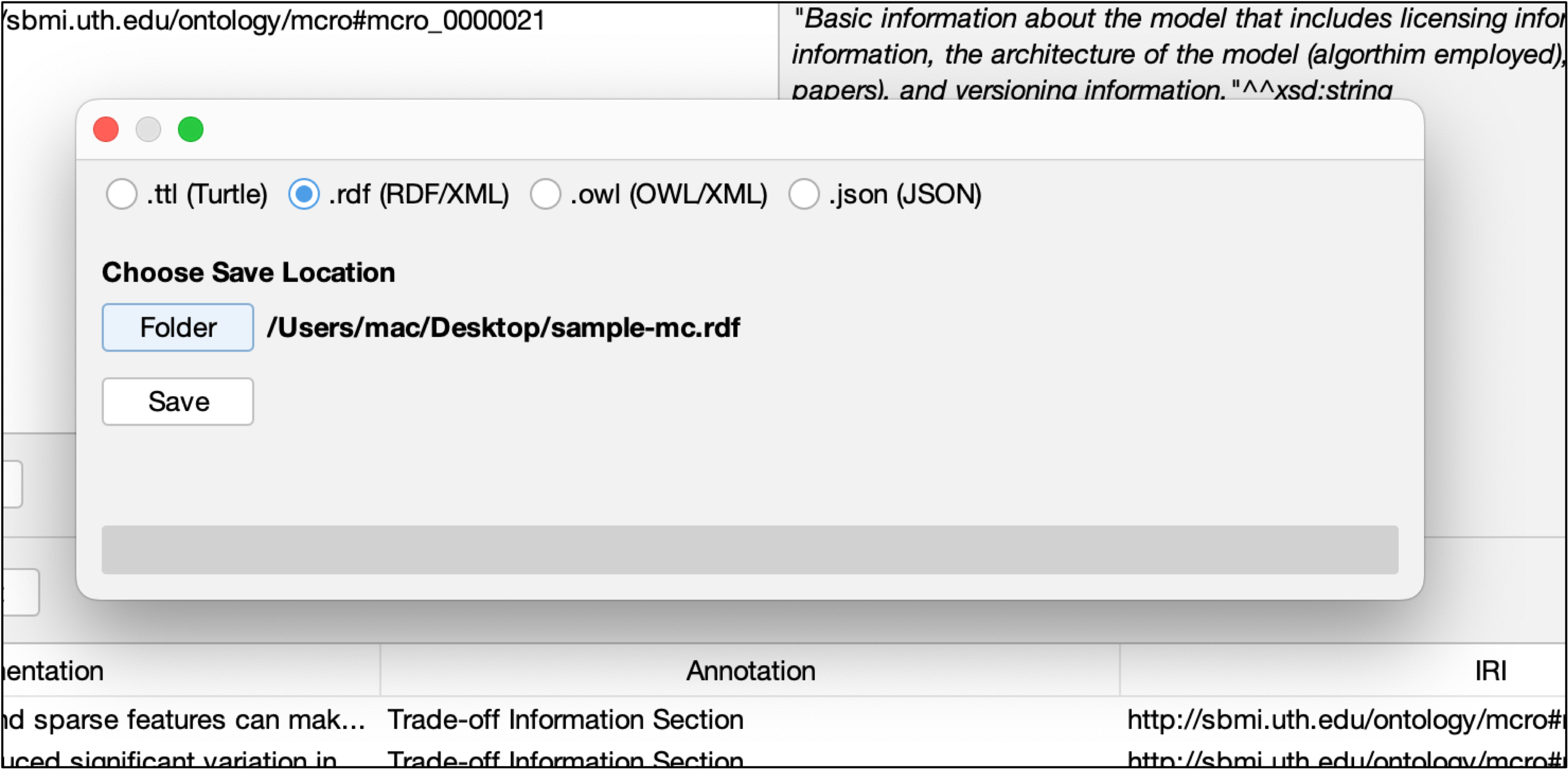}
	\caption{A user interface prompt for options to export the final ontology-based model card.}
	\label{fig:export-app}
\end{figure}

Clicking "Export" (see Figure \ref{fig:collection-app}) calls on the software library to link the textual information with the associated MCRO classes, and then adds it to the MCRO schema. The library then generates the ontology-based model card (see Algorithm \ref{alg:dl}) and also provides options to export in different computable formats - Turtle, RDF, OWL, and JSON. Figure \ref{fig:export-app} shows the prompt for a prospective user to choose the desired format.

\FloatBarrier
\section{Discussion}

The goal of model card reports is to communicate a transparent description of machine learning-based AI models that includes information about utilized datasets, limitations, intended use, etc. However, when the reports are published as static, non-computable documents on the web, it presents some barriers that prevent advanced capabilities such as querying and machine reasoning on data and information about machine learning models used in biomedical sciences. Additionally, authoring an ontology-based model card report through Prot\'eg\'e or a text editor poses some inconveniences to have a computable and linkable model card report. What our early work shows, in addition to having the capacity to generate machine-processable model cards, is we could possibly develop tools using ontology-driven and semantic web backend application services to publish ontology-based model card reports. 

\begin{figure}[h!]
	\centering
	\includegraphics[width=1.1\linewidth]{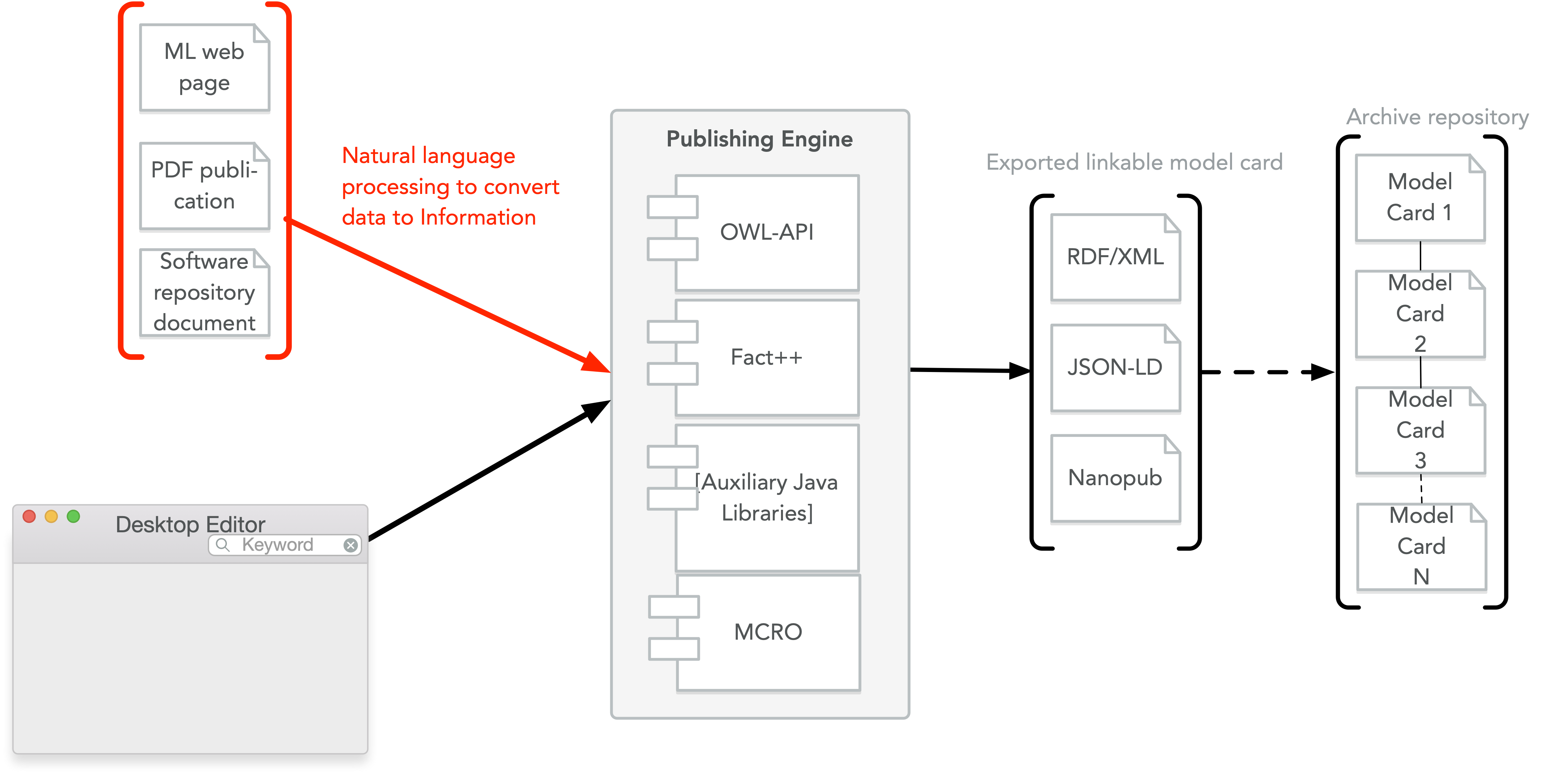}
	\caption{Generalized description of the publishing pipeline for linked model card reports. Red highlights the future step to address.}
	\label{fig:pipeline-future}
\end{figure}

While this work is an incremental step, we envision a more comprehensive system to support a pipeline and publishing ecosystem for ontology-based model card reports. Figure \ref{fig:pipeline-future} shows how this work complements the system and also other components to fully realize in future work. 

Natural language processing aims to extract unstructured (and semi-structured) text from data (i.e., data to structured information). One possibility for us to investigate is to utilize natural language processing (NLP) to automatically (or semi-automatically) produce the text information of the machine learning model.  The software publishing library will import the NLP-generated information and proceed with the same process as described in this paper. As an AI task, NLP methods can extract information from static resources (web pages that host machine learning models, research papers about machine learning models, text data from software repositories, etc.) and feed it to the ontology-driven engine.

Aside from the NLP method for model card information extraction, we also plan to reevaluate and develop an editor interface for the publishing engine that is different from what is presented in this paper. The front end application was intended to show how atomic textual information with only an annotation from MCRO can be linked with reasoning based off the MCRO model. Alongside with an offline desktop version, other possibilities include a web-based editor that could support archiving. A usability assessment will be considered with potential targeted end users.

\section{Conclusion}
 In this paper, we describe how an ontology that we developed from our previous study can operationalize the publishing of computable, linkable model card report. The output contrasts from a static document that has less processing characteristics for systems to utilize (e.g., archiving and querying). Our planned architecture utilizes semantic web technology, Description Logic, and experimental software to produce a preliminary proof of concept publishing engine to generate semantic-based model card reports to support FAIR principles. While we have demonstrated the feasibility of our ontology-based approach there are a few future steps to realize a serviceable model card publishing engine.
\FloatBarrier
\begin{acks}
 This research was supported by NIH grants under Award Numbers RF1AG072799, R56AI150272, R56AG074604, and T15LM007093, and by the Cancer Prevention Research Institute of Texas under Award Number \#RP220244.
\end{acks}

\bibliographystyle{ACM-Reference-Format}
\bibliography{references}

%
%
%
%
%
%
%
%

\end{document}